\documentclass[aps,prl,twocolumn,showpacs,groupedaddress]{revtex4}
\usepackage{graphicx}

\begin{document}


\title{Cold Molecule Spectroscopy for Constraining the Evolution of the Fine Structure Constant}


\author{Eric R. Hudson}
\email[]{ehudson@jilau1.colorado.edu}
\author{H. J. Lewandowski}
\author{Brian C. Sawyer}
\author{Jun Ye}
\affiliation{JILA, National Institute of Standards and Technology
and University of Colorado \\ Department of Physics, University of
Colorado, Boulder, CO 80309-0440}


\date{\today}

\begin{abstract}
We report precise measurements of ground-state, $\lambda$-doublet
microwave transitions in the hydroxyl radical molecule (OH).
Utilizing slow, cold molecules produced by a Stark decelerator we
have improved over the precision of the previous best measurement by
twenty-five-fold for the F' = 2 $\rightarrow$ F = 2 transition,
yielding (1 667 358 996 $\pm$ 4) Hz, and by ten-fold for the F' = 1
$\rightarrow$ F = 1 transition, yielding (1 665 401 803 $\pm$ 12)
Hz. Comparing these laboratory frequencies to those from OH
megamasers in interstellar space will allow a sensitivity of 1 ppm
for $\Delta\alpha/\alpha$ over $\sim$$10^{10}$ years.
\end{abstract}


\maketitle


Current theories that attempt to unify gravity with the other
fundamental forces predict spatial and temporal variations in the
fundamental constants, including the fine structure constant,
$\alpha$ \cite{Olive:2004}. Measurements of the variation of
$\alpha$ by observation of multiple absorption lines from distant
quasars are currently not in agreement \cite{Webb:2001,Quast:2004}.
Due to the use of spatially diverse absorbers, these measurements
are sensitive to relative Doppler shifts. Therefore an independent
confirmation of the variation of $\alpha$ is important. Recently,
there has been much interest in using OH megamasers in interstellar
space to constrain the evolution of fundamental constants
\cite{Darling:2003,Chengalur:2003,Kanekar:2004} with several
important advantages. Specifically, it has been shown that the sum
and difference of the $\Delta$F = 0 (F is total angular momentum)
transition frequencies in the ground $\lambda$-doublet of OH depend
on $\alpha$ as $\alpha^{0.4}$ and $\alpha^{4}$, respectively
\cite{Darling:2003}. Thus, by comparing the values measured from OH
megamasers to laboratory values it is possible to constrain $\alpha$
over cosmological time. Most importantly, the multiple lines (that
have different dependence on $\alpha$) arising from a single
localized source differentiate the relative Doppler shift from the
$\Delta\alpha/\alpha$ measurement. Furthermore, because of the
unique properties of the $\lambda$-doublet, the $\Delta$F = 0
transitions are extremely insensitive to magnetic fields. However,
as pointed out by Darling \cite{Darling:2003}, for the current
limits on $\Delta\alpha/\alpha$ the change in the relevant
measurable quantities is on the order of 100 Hz, which prior to this
work was the accuracy of the best laboratory based measurement.
Thus, Darling called on the community to produce a more precise
measurement of the OH $\lambda$-doublet microwave transitions to
allow for tighter constraints on $\Delta\alpha/\alpha$.

Despite the prominence of the OH radical in molecular physics, the
previous best measurement of the OH ground $\lambda$-doublet,
performed by ter Meulen and Dymanus \cite{TerMeulen:1971}, has stood
for over 30 years. This lack of improvement was due to the
relatively slow progress in the center-of-mass motion control of
molecules, which limited the maximum field interrogation time, and
thus the spectroscopic resolution. The ability of a Stark
decelerator \cite{Bethlem:1999,Bochinski:2003} to provide slow, cold
pulses of molecules makes it an ideal source for molecular
spectroscopy \cite{VanVeldhoven:2004}. In this work, a Stark
decelerator is used along with standard microwave spectroscopy
techniques to perform the best measurement to date of the $\Delta$F
= 0, $\lambda$-doublet microwave transitions in OH, which along with
appropriate astrophysical measurements can be used to constrain
$\Delta\alpha/\alpha$ with a sensitivity of 1 ppm over the last
$\sim$$10^{10}$ years.

\begin{figure}
\resizebox{1\columnwidth}{!}{
    \includegraphics{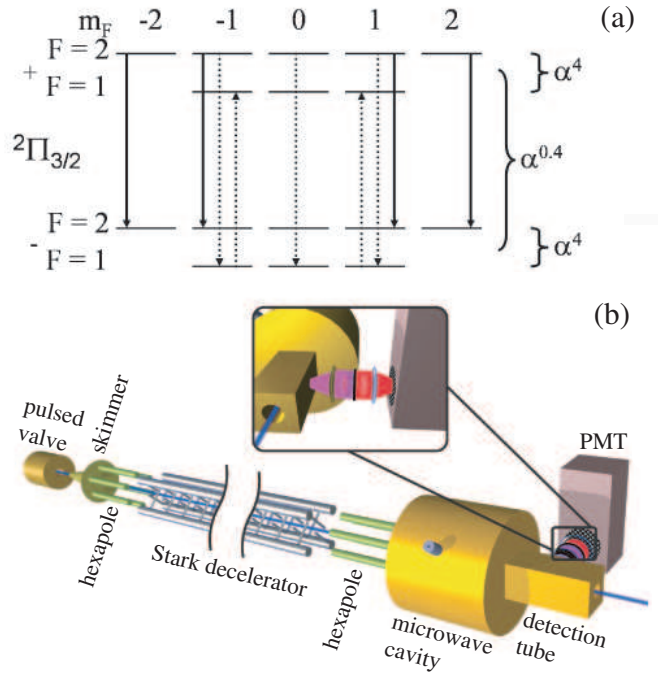}%
} \caption{(Color online) a) OH ground $\lambda$-doublet state. The
arrows represent the effect of the applied microwave pulses for the
2$\rightarrow$2 (solid arrows) and 1$\rightarrow$1 (dotted arrows)
transitions. b) Schematic of experiment (inset depicts detection
region). \label{Figure1}}
\end{figure}

In its ro-vibronic ground state OH is a Hund's case (a) molecule
with a $^2\Pi$ configuration and total molecule-fixed angular
momentum of $\Omega = \frac{3}{2}$. For the most abundant isotopomer
(O$^{16}$H) the oxygen has no nuclear spin and the hydrogen carries
a nuclear spin of $\frac{1}{2}$ leading to two total spin states
with F = 1 and 2. Because the unpaired electron in OH has one unit
of orbital angular momentum, these ro-vibronic ground states are
`$\lambda$-doubled', leading to the closely spaced opposite parity
$\lambda$-doublet states shown in Fig. \ref{Figure1}(a) labeled as
$|F, m_F, \rm{parity}\rangle$. Though molecules are decelerated only
in the $|2,\pm2,+\rangle$ and $|2,\pm1,+\rangle$ states, the
field-free region from the hexapole to the microwave cavity leads to
an equal redistribution of the population among the five magnetic
sub-levels of the F = 2 upper doublet state. In this work, both
$\Delta$F = 0 electric dipole transitions between the upper and
lower doublet states are studied. Though these transitions exhibit
large Stark shifts (as is necessary for Stark deceleration), they
show remarkably small Zeeman shifts. In fact, because the magnetic
dipole operator respects parity, one expects the $\Delta$F = 0
transitions of a pure case (a) molecule to show no magnetic field
dependence. Nonetheless, because the hyperfine splitting differs in
the upper and lower doublet, and part of the $|m_F|$ = 1 magnetic
dipole moment comes from mixing with the other hyperfine component,
there is a small quadratic shift of the $|2,\pm1,+\rangle$
$\rightarrow$ $|2,\pm1,-\rangle$ transition frequency (150
Hz/Gauss$^2$). Furthermore, because OH is not completely case (a)
(\textit{i.e.} the electron's orbital angular momentum and spin are
slightly decoupled from the axis) the g-factor is slightly larger in
the lower doublet \cite{Radford}. Thus, transitions between the $m_F
>$ 0 ($m_F <$ 0) components are blue-shifted (red-shifted) relative
to the zero field value. Hence for experiments probing both positive
and negative $m_F$ components such as this work, the g-factor
difference leads to a broadening and eventually a bifurcation in the
lineshape for increasing magnetic field. This effect, explored at
relatively large fields (0.6 - 0.9 T), was found to shift the
$\Delta$F = 0 transition frequencies of the $|2,\pm2,+\rangle,
|2,-1,+\rangle$, and $|1,1,+\rangle$ states at a rate of
sign$(m_F)\times2.7$ kHz/Gauss, and the $|1,-1,+\rangle$ and
$|2,1,+\rangle$ states at a rate of sign$(m_F)\times0.9$ kHz/Gauss
\cite{Radford}. For even population distribution this effect leads
to a shift of -450 Hz/Gauss and 900 Hz/Gauss for the 2$\rightarrow$2
and 1$\rightarrow$1 transitions, respectively.

\begin{figure}
\resizebox{1\columnwidth}{!}{
    \includegraphics{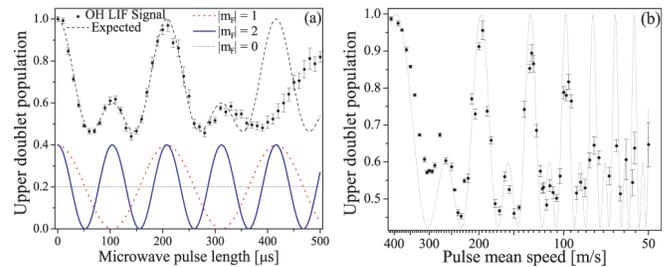}%
} \caption{(Color online) a) Rabi-flopping as a function of time.
The contributions of the individual magnetic sub-levels are shown
near the bottom, while the sum is shown as the dashed line. b)
Rabi-flopping as a function of velocity for a fixed (spatial) length
microwave pulse. The dashed line is the expected. \label{RabiFlop}}
\end{figure}

Shown in Fig. \ref{Figure1}(b) is a schematic of our experimental
setup. OH molecules seeded in Xenon are created in a pulsed
discharge \cite{Lewandowski:2004} producing a molecular pulse with a
mean speed of 410 m/s and 10\% longitudinal velocity spread.
Following a skimmer is an electrostatic hexapole used to
transversely couple the molecules into our Stark decelerator, which
is described in detail elsewhere
\cite{Bochinski:2003,Bochinski:2004,Hudson:2004}. The Stark
decelerator is used in a variety of operating conditions, yielding
pulses of molecules at a density of 10$^{6}$ cm$^{-3}$ with mean
speeds chosen between 410 m/s and 50 m/s with longitudinal
temperatures between 1 K and 5 mK, respectively. Once the molecular
pulse exits the decelerator it is focused by a second hexapole to
the detection region. Located between the second hexapole and the
detection region is a 10 cm long cylindrical microwave cavity with
its axis aligned to the molecular beam. The microwave cavity is
operated near the TM$_{010}$ mode, such that the electric field is
extremely uniform over the region sampled by the molecules
\cite{Jackson}. The microwave cavity is surrounded by highly
magnetically permeable material to provide shielding from stray
magnetic fields. From measurements of the magnetically sensitive F =
2$\rightarrow$F = 1 transition frequency the residual field in the
cavity was determined to be $<$ 2 milliGauss. This small field
results in an absolute Zeeman shift of $<$ 0.9 Hz and $<$ 1.8 Hz for
the 2$\rightarrow$2 and 1$\rightarrow$1 transitions, respectively.
Note that this is a cautious upper-bound on the Zeeman shift, since
it is based on work performed at $\sim$10$^7$ larger magnetic field
where the angular momentum decoupling is enhanced. Furthermore, it
was verified within experimental resolution that there were no
transition frequency shifts due to stray electric fields.

Two independently switchable microwave synthesizers, both referenced
to a Cesium standard, are used to provide the microwave radiation
for driving the $\lambda$-doublet transitions. For probing the
$|2,m_F,+\rangle$ $\rightarrow$ $|2,m_F,-\rangle$ $\lambda$-doublet
transition (2$\rightarrow$2) only one synthesizer is needed since
the molecules enter the cavity in the F = 2 state. Accordingly, two
synthesizers are needed for probing the 1$\rightarrow$1 transition.
The first synthesizer transfers molecules from the F = 2 upper
doublet state to the F = 1 lower doublet state. The second
synthesizer then probes the 1$\rightarrow$1 transition. After the
molecules exit the cavity the population of the upper doublet is
probed by laser-induced fluorescence (LIF). For the LIF measurement
the molecules are excited along the $^2\Sigma_{1/2} (v = 1)
\leftarrow$ $^2\Pi_{3/2} (v = 0)$ line at 282 nm by light produced
from a doubled pulsed-dye laser. This excited state decays primarily
along $^2\Sigma_{1/2} (v = 1) \rightarrow$ $^2\Pi_{3/2} (v = 1)$ at
313 nm. The red-shifted fluorescence is collected and imaged onto a
photomultiplier tube coupled to a multi-channel scaler.

To characterize the performance of the microwave cavity the
population in the upper doublet was recorded as a function of
applied microwave pulse length as shown in Fig. \ref{RabiFlop}(a)
(so-called Rabi-flopping). For the data shown, the molecules were
decelerated to 200 m/s and a microwave pulse resonant with the
2$\rightarrow$2 transition was applied such that its midpoint time
coincided with the molecules being at the cavity center. Thus, as
the pulse length was increased the molecules encountered a pulse
that grew symmetrically about the cavity center. Because the LIF
detection scheme is sensitive to all molecules in the upper doublet
and the applied microwave field simultaneously drives transitions
between the different magnetic sub-levels, the Rabi-flopping signal
is more complicated than the traditional
$\sin^2(\frac{\omega'_Rt_p}{2})$. Here $\omega'_R$ is the effective
Rabi frequency given in terms of the detuning, $\delta$, and Rabi
frequency, $\omega_R$, as $\omega'_R = \sqrt{\delta^2 +
\omega_R^2}$, and $t_p$ is the microwave pulse length. As shown in
Fig. \ref{Figure1}(a) for the 2$\rightarrow$2 transition (solid
arrows), both the $|m_F|$ = 2 and $|m_F|$ = 1 magnetic sub-levels
are driven by the microwave field (the $|m_F|$ = 0 level has a zero
transition moment). Because the electric dipole transition moment of
the $|m_F|$ = 2 is twice that of $|m_F|$ = 1 \cite{Avdeenkov:2002},
the Rabi-flopping signal exhibits beating. The calculated individual
magnetic sub-level contributions are shown at the bottom of Fig.
\ref{RabiFlop}(a) with their sum represented by the dashed line
plotted over the data. The data points were determined by comparing
the populations in the upper doublet at the detection region with
and without the microwave field applied. Clearly, the behavior is
exactly as expected until $t_p \geq$ 300 $\mu$s when $\omega_R$ of
both the $|m_F|$ = 2 and $|m_F|$ = 1 transitions appears to
decrease. This reduction in $\omega_R$ is the result of the electric
field diminishing near the cavity end-caps \cite{Efieldnote}.

\begin{figure}
\resizebox{1\columnwidth}{!}{
    \includegraphics{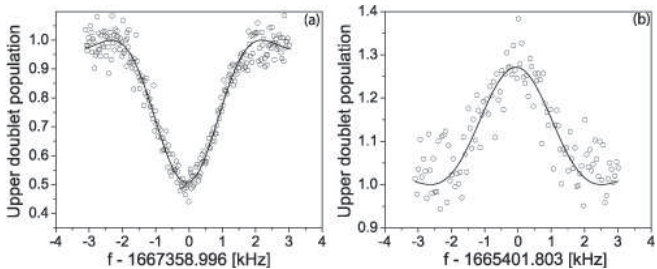}%
} \caption{Representative line shape of the 2$\rightarrow$2 (a) and
1$\rightarrow$1 (b) transitions. Both measurements correspond to an
interaction time of 0.5 ms (v = 200 m/s). A center frequency is
extracted from each fit (solid line). \label{LineGraph}}
\end{figure}

Alternatively to fixing the molecular velocity, $v$, and varying
$t_p$, the Stark decelerator allows $v$ to be varied while applying
a fixed (spatial) length microwave pulse. This allows a check of
systematics associated with beam velocity. Data taken in this manner
is shown in Fig. \ref{RabiFlop}b, where a microwave pulse resonant
with the 2$\rightarrow$2 transition was applied for the entire time
the molecules were in the cavity. The microwave power for this
measurement was chosen such that molecules with a 400 m/s velocity
underwent one complete population oscillation (\textit{i.e.} a
2$\pi$ pulse for the $|m_F| = 1$ and a 4$\pi$ pulse for the $|m_F|$
= 2 transitions). This was done so that population revivals occurred
at velocities that were integer sub-multiples of 400 m/s. While the
behavior agrees well with the expected (dotted line), there are two
noticeable deviations. First, for 270 m/s $\leq v \leq$ 400 m/s the
fringe visibility is less than expected. This is because, as
detailed in our earlier work \cite{Bochinski:2004,Hudson:2004},
molecules with these relatively high velocities have not been
decelerated out of the background molecular pulse. Thus, molecules
with a large distribution of speeds (as compared to the decelerated
molecules) are detected, leading to reduced contrast. Second,
substantial decoherence is observed for $v \leq$ 130 m/s. The source
of this decoherence has been experimentally determined as the result
of microwave radiation leaking from the cavity and being reflected
off the decelerator back into the cavity. It is interesting to note
that since the metal detection tube acts as a waveguide with a
cut-off frequency much higher than that applied, no radiation leaks
from the rear of the cavity. Thus, future experiments should include
a small waveguide section on both sides of the cavity to prevent
leakage. Furthermore, because this decoherence is due to reflected
radiation it is microwave power dependent, and presents no problem
for the transitions frequency measurements, which use $<$ 10\% of
the microwave power used in Fig. \ref{RabiFlop}, such that no
noticeable decoherence occurs.

Because two magnetic sub-levels with differing $\omega_R$'s undergo
the 2$\rightarrow$2 transition, the traditional Rabi-spectroscopy
method of applying a $\pi$ pulse over the length of the cavity is
not optimal. Thus, for measurements of the 2$\rightarrow$2
transition frequency the microwave power was chosen such that for
$\delta$ = 0 the maximum contrast was produced with as small a
microwave power as possible (\textit{i.e.} transfer the molecules to
the first dip in Fig. \ref{RabiFlop}(a) over the length of the
cavity). Representative data for the 2$\rightarrow$2 transition is
shown in Fig. \ref{LineGraph}(a) where the driving frequency, f, is
varied under a fixed microwave power. Data points in this graph were
generated by comparing the population in the upper doublet at the
detection region with and without the probing microwave field. The
fit to the data (solid line) is generated from the typical Rabi
lineshape formula except contributions from the different magnetic
sub-levels are included. For probing the 1$\rightarrow$1 transition
it is necessary to prepare the molecules in a F = 1 level since they
originate in the $|2,m_F,+\rangle$ level from the Stark decelerator.
This is accomplished by using the first 70 $\mu$s the molecules
spend in the cavity to drive them on the satellite 2$\rightarrow$1
line, yielding molecules in the $|1,\pm1,-\rangle$ and
$|1,0,-\rangle$ states (downward dotted arrows in Fig.
\ref{Figure1}). The remaining time the molecules spend in the cavity
(depending on $v$) is then used to probe the 1$\rightarrow$1
transition frequency by applying a $\pi$ pulse, which transfers only
the $|m_F|$ = 1 molecules to the upper doublet (upward dotted arrows
in Fig. \ref{Figure1}). Representative data for the 1$\rightarrow$1
transition is shown in Fig. \ref{LineGraph}b. In contrast to the
2$\rightarrow$2, line the LIF signal is maximum on resonance because
molecules are being transferred into the detected upper doublet.
Also, because only 40\% of the population participate in the
transitions (\textit{i.e.} the $|m_F| = 1$) the contrast between on
\begin{figure}
\resizebox{1\columnwidth}{!}{
    \includegraphics{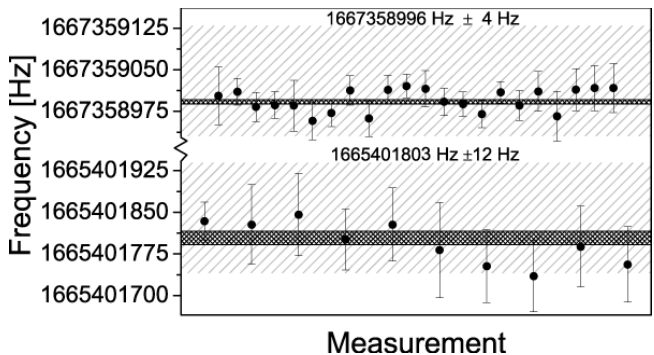}%
} \caption{Results of multiple measurements of the 2$\rightarrow$2
transition (upper) and the 1$\rightarrow$1 transition (lower).
\label{FreqCum}}
\end{figure}and off resonance is reduced relative to the 2$\rightarrow$2
transition, leading to a slightly larger error for determining the
center frequency. For both panels of Fig. \ref{LineGraph} the
molecules were decelerated to 200 m/s yielding linewidths of 2 kHz.
A typical fit determines the center frequency within 20 to 50 Hz,
depending on the transition. The results of several measurements of
both the 2$\rightarrow$2 and 1$\rightarrow$1 center frequencies are
displayed in Fig. \ref{FreqCum}. Each point and its error bar
represents the result of a fit to a measured lineshape like those
shown in Fig. \ref{LineGraph}. Using the standard error of each fit
as a weight the mean and standard error of the transition
frequencies are found to be (1 667 358 996 $\pm$ 4) Hz and (1 665
401 803 $\pm$ 12) Hz for the 2$\rightarrow$2 and 1$\rightarrow$1,
respectively. For comparison, the lightly hatched boxes represent
the bounds set on the transition frequency by the previous best
measurement \cite{TerMeulen:1971}, while the limits produced by this
measurement are displayed as darker cross-hatched boxes. The
transition frequencies reported here are limited only by statistical
uncertainties.

\begin{figure}
\resizebox{1\columnwidth}{!}{
    \includegraphics{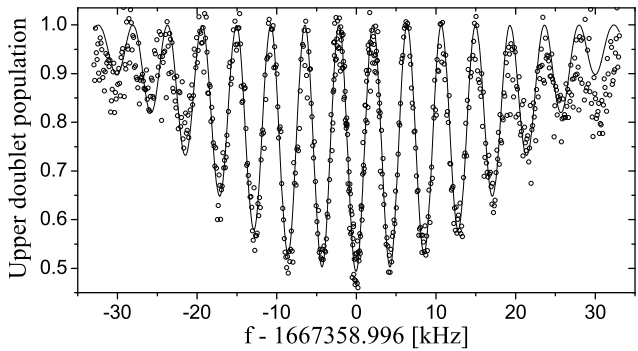}%
} \caption{Ramsey spectroscopy for the 2$\rightarrow$2 transition
with 0.2 ms pulse separation time. \label{Ramsey}}
\end{figure}

The Ramsey technique of separated pulses inside the same microwave
cavity was also used to measure the transition frequencies as seen
in Fig. \ref{Ramsey} with a resolution comparable to the reported
Rabi measurements. However, because the molecular pulses are
extremely monochromatic, minimal gain was observed in the recovered
signal-to-noise ratio. This technique will of course be critical for
any future molecular fountain clock. The molecular clock could enjoy
reduced systematic shifts, such as the magnetic field insensitive
transitions demonstrated in this work. It will be most interesting
to compare an atomic clock against a molecular one which depends
differently on the fine structure constant.

In summary, microwave spectroscopy was performed on slow, cold
molecular pulses produced by a Stark decelerator resulting in the
most precise measurement of the OH $\Delta$F = 0, $\lambda$-doublet
transitions. These results along with appropriate astrophysical
measurement of OH megamasers can be used to produce constraints on
$\Delta\alpha/\alpha$ with a sensitivity of 1 ppm over the last
$\sim$$10^{10}$ yr. At the same time the use of cold molecules for
the most precise molecular spectroscopy has been demonstrated.
Specifically, by producing slow, cold molecular packets the Stark
decelerator allows increased interrogation time, while virtually
eliminating any velocity broadening.

\begin{acknowledgments}
The authors are indebted to Steven Jefferts and John L. Hall for
crucial contributions. This work is supported by NSF, NIST, DOE, and
the Keck Foundation.
\end{acknowledgments}

\bibliography{OH_Microwave_Bib}

\begin{thebibliography}{17}
\bibitem[{Efi()}]{Efieldnote}
\bibinfo{note}{For the TM$_{010}$ mode the electric field is uniform along the
  cavity axis. However, due to the use of multiple frequencies, we work below
  the TM$_{010}$ cavity resonance leading to the observed electric field
  decay.}
\end{thebibliography}

\end{document}